\documentclass[sn-mathphys-num]{sn-jnl}

\usepackage{graphicx}%
\usepackage{multirow}%
\usepackage{amsmath,amssymb,amsfonts}%
\usepackage{amsthm}%
\usepackage{mathrsfs}%
\usepackage[title]{appendix}%
\usepackage{xcolor}%
\usepackage{textcomp}%
\usepackage{manyfoot}%
\usepackage{booktabs}%
\usepackage{algorithm}%
\usepackage{algorithmicx}%
\usepackage{algpseudocode}%
\usepackage{listings}%

\theoremstyle{thmstyleone}%
\theoremstyle{thmstyletwo}%

\theoremstyle{thmstylethree}%

\begin{document}

\title[Article Title]{A Framework for Predictive Directional Trading Based on Volatility and Causal Inference}


\author*[1]{\fnm{Ivan} \sur{Letteri}}\email{ivan.letteri@univaq.it}


\affil*[1]{\orgdiv{Department of Life, Health and Environmental Sciences}, \orgname{University of L’Aquila}, \orgaddress{\street{P.le S. Tommasi}, \city{Coppito}, \postcode{67100}, \state{L'Aquila}, \country{Italy}}}




\abstract{\textbf{Purpose:} This study introduces a novel framework for identifying and exploiting predictive lead-lag relationships in financial markets. We propose an integrated approach that combines advanced statistical methodologies with machine learning models to enhance the identification and exploitation of predictive relationships between equities.\\
\textbf{Methods:} We employed a Gaussian Mixture Model (GMM) to cluster nine prominent stocks based on their mid-range historical volatility profiles over a three-year period. From the resulting clusters, we constructed a multi-stage causal inference pipeline, incorporating the Granger Causality Test (GCT), a customised Peter-Clark Momentary Conditional Independence (PCMCI) test, and Effective Transfer Entropy (ETE) to identify robust, predictive linkages. Subsequently, Dynamic Time Warping (DTW) and a K-Nearest Neighbours (KNN) classifier were utilised to determine the optimal time lag for trade execution. The resulting strategy was rigorously backtested.\\
\textbf{Results:} The proposed volatility-based trading strategy, tested from 8 June 2023 to 12 August 2023, demonstrated substantial efficacy. The portfolio yielded a total return of 15.38\%, significantly outperforming the 10.39\% return of a comparative Buy-and-Hold strategy. Key performance metrics, including a Sharpe Ratio up to 2.17 and a win rate up to 100\% for certain pairs, confirmed the strategy's viability.\\
\textbf{Conclusion:} This research contributes a systematic and robust methodology for identifying profitable trading opportunities derived from volatility-based causal relationships. The findings have significant implications for both academic research in financial modelling and the practical application of algorithmic trading, offering a structured approach to developing resilient, data-driven strategies.}




\maketitle

\section{Introduction}\label{sec1}
The financial sector is increasingly adopting volatility-based trading strategies, capitalising on inherent market dynamics to optimise returns. Within this context, Artificial Intelligence (AI) has emerged as a pivotal technology, equipping traders with advanced tools to analyse and exploit market volatility. The capacity of AI to estimate mean volatility offers crucial insights into the uncertainties and risks associated with individual securities and broader market trends \cite{letteri2022dnnforwardtesting}.

Accurate intraday volatility forecasts are essential for effective risk management \cite{jjfinecnbr005}, enabling traders to assess potential price fluctuations, particularly in AI-driven automated trading systems \cite{bates2019}. The integration of statistical and machine learning techniques has profoundly transformed the development of profitable trading strategies in recent years \cite{MLPletteriStockTrading}.

Building upon a preliminary study \cite{femibLetteri24} presented at the FEMIB 2024 conference\footnote{FEMIB: Finance, Economics, Management, and IT Business, FEMIB 2024, Angers, France, 28-29 April 2024}, this article explores the efficacy of the Gaussian Mixture Model (GMM) algorithm \cite{gmm2014} in analysing the mid-range volatility of major stocks. This research seeks to examine inter-stock relationships by analysing volatility patterns and employing causal inference algorithms to identify predictive signals for directional trading strategies. By combining these advanced techniques, the study aims to classify stocks with similar volatility behaviours and uncover predictive relationships—such as lead-lag effects, which are conceptually distinct from but can be as informative as cointegration—that can inform directional trading strategies \cite{cointegration2013}. This approach enhances predictive accuracy and facilitates more informed decision-making in financial markets.

Our work is driven by the following research questions: 
\begin{itemize}
    \item $RQ_1$: Can Gaussian Mixture Models effectively classify stocks into distinct clusters based on their mid-range volatility profiles? 
    \item $RQ_2$: For stocks exhibiting similar mid-term volatility, can a multi-stage causal inference approach, combining the Granger Causality Test (GCT) and a customised Peter-Clark Momentary Conditional Independence (PCMCI) test, effectively identify directional predictive influences among them? 
    \item $RQ_3$: Once a subset of influential stock pairs is identified, can a combination of Dynamic Time Warping (DTW), Transfer Entropy (TE), and K-Nearest Neighbours (KNN) quantify the temporal delay between them to formulate a profitable trading strategy?
\end{itemize}

By addressing these questions, we aim to develop a robust trading system that utilises influential stocks as lead indicators. Through rigorous backtesting and performance analysis, we validate the reliability of the proposed lead-lag trading strategy, offering insights into its efficiency in identifying profitable opportunities within the complexities of financial markets.

Our prior work on technical trading strategies emphasised the use of technical indicators for timing investments \cite{letteri2022dnnforwardtesting, LetteriFEMIB23}. In this paper, we shift our focus to historical volatility estimators as the primary data source for stock selection. We employ a causal inference framework based on the Granger Causality Test \cite{Engle1987}, complemented by Effective Transfer Entropy (ETE) and a customised PCMCI test, to analyse asset price dynamics in depth. This methodology is integrated into the Volatility Trading System (Vol-TS) module of the AITA framework \cite{letteri2023volts, ital-ia_Letteri24}, which classifies securities based on mid-term volatility and identifies predictive price relationships among them. Vol-TS is a component of a larger architecture involving Green Economy and Generative AI activities presented at the Ital-IA 2025 conference \footnote{\texttt{Ital-IA:} \url{https://www.ital-ia2025.it/workshop/ai-generativa.html}, last access July 2025}.

This paper is organised as follows: Section \ref{sect:prel} introduces foundational concepts related to the AITA framework, focusing on the Vol-TS module and the Aita BackTesting (Aita-BT) engine. Section \ref{sect:methodology} outlines the methods for volatility analysis and causal inference. Section \ref{sect:results} presents the empirical findings from our backtesting experiments, including a robustness analysis. Section \ref{sect:discussion} interprets these results and compares them with existing literature. Finally, Section \ref{sect:conclusion} summarises the contributions of this work and suggests directions for future research.
 
\section{Background}\label{sect:prel}
\subsection{Price Action and Volatility}
\label{sec:techAnalysis}
Price action (PA) directly influences historical volatility (HV), which in turn can offer valuable insights into future PA. HV typically increases when PA exhibits strong directional movements, such as breakouts, or wide trading ranges. Conversely, low HV often indicates a period of consolidation, potentially signalling an impending increase in volatility.

The Vol-TS module, a component of the AITA framework, operates on these principles. Within Vol-TS, PA is encoded using the Open, High, Low, Close (OHLC) price format. For each timeframe $t$, the OHLC of an asset is represented as a four-dimensional vector $X_t = (x^{(o)}_t, x^{(h)}_t, x^{(l)}_t, x^{(c)}_t)^T$, where $x_t^{(l)} > 0$, $x_t^{(l)} \le x_t^{(h)}$, and $x_t^{(o)}, x_t^{(c)} \in [x_t^{(l)}, x_t^{(h)}]$.

\begin{figure}[!ht]
	\centerline{\includegraphics[width=36em]{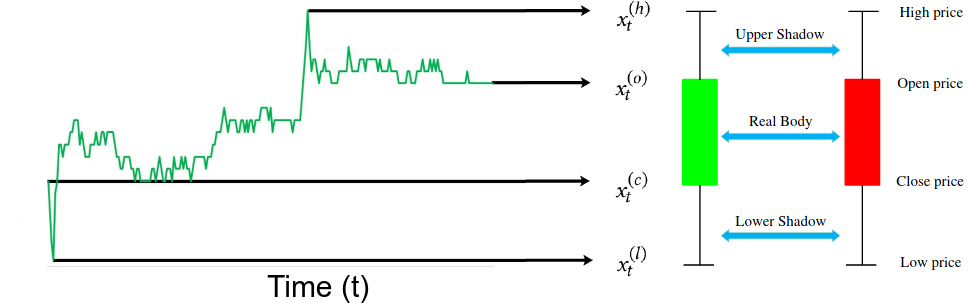}}
	\caption{Example of a candlestick chart. \texttt{This figure is from Figure 1 in Letteri \cite{femibLetteri24}.}}
	\label{fig:candlestick}
\end{figure}

The module integrates advanced data analytics and machine learning to enhance its predictive capabilities. By clustering assets based on volatility patterns and employing causal inference tests, it identifies predictive relationships between stocks. This enables traders not only to anticipate the movement of individual stocks but also to comprehend how different assets influence one another, thereby providing a more holistic perspective of market dynamics.

\subsection{Historical Volatility Estimators}
\label{sect:hvs}
The Historical Volatility (HV) module is central to our strategy, employing a range of estimators to capture distinct aspects of price fluctuations from OHLC data. We utilise estimators that incorporate intra-day price movements, offering a more nuanced view than the simple standard deviation of closing prices. The high-quality OHLC data is sourced via the AITA framework's integration with MetaTrader5\footnote{\texttt{MT5:} \url{https://www.metatrader5.com/}, last access Aug 2024}.

\begin{itemize}
    \item[-] The \textit{Parkinson} (PK) estimator incorporates the daily high and low prices: 
    $$ \sigma_{PK}^2 = \frac{1}{4N\ln(2)}\sum_{t=1}^N \left(\ln\frac{x_t^{(h)}}{x_t^{(l)}}\right)^2 $$
    It is particularly effective at capturing volatility during trending periods.
    
    \item[-] The \textit{Garman-Klass} (GK) estimator extends the PK estimator by including opening and closing prices: 
    $$ \sigma_{GK}^2 = \frac{1}{N}\sum_{t=1}^N \left[ \frac{1}{2}\left(\ln\frac{x_t^{(h)}}{x_t^{(l)}}\right)^2 - (2\ln(2)-1)\left(\ln\frac{x_t^{(c)}}{x_t^{(o)}}\right)^2 \right] $$
    This estimator is designed to be more efficient by using more information from the trading day.
    
    \item[-] The \textit{Rogers-Satchell} (RS) estimator accounts for price drift and is independent of the opening price gap: 
    $$ \sigma_{RS}^2 = \frac{1}{N}\sum_{t=1}^{N} \left[ \ln\left(\frac{x_t^{(h)}}{x_t^{(c)}}\right)\ln\left(\frac{x_t^{(h)}}{x_t^{(o)}}\right) + \ln\left(\frac{x_t^{(l)}}{x_t^{(c)}}\right)\ln\left(\frac{x_t^{(l)}}{x_t^{(o)}}\right) \right] $$
    
    \item[-] The \textit{Yang-Zhang} (YZ) estimator \cite{Yang2000Zhang} is a weighted average of the RS estimator, the closing price volatility, and the overnight price gap volatility, aiming for minimum variance: 
    $$ \sigma_{YZ}^2 = \sigma_{o}^2 + k\sigma_{c}^2 + (1-k)\sigma_{RS}^2 $$
    where $\sigma_{o}^2$ is the overnight volatility, $\sigma_{c}^2$ is the close-to-open volatility, $\sigma_{RS}^2$ is the Rogers-Satchell variance, and $k = \frac{0.34}{1.34 + (N+1)/(N-1)}$. The YZ estimator is considered one of the most robust, particularly in markets with jumps.
\end{itemize}

In this research, our focus is on mid-range volatility. This allows us to avoid trading when volatility is excessively high, mitigating risk, or too low, offering limited profit potential. These advanced estimators provide a more comprehensive understanding of volatility compared to standard deviation, adapting to various market conditions and better handling the full spectrum of daily price information.

\subsection{Trading Strategies}
\label{subsect:tsu}
The AITA framework supports several trading strategies. For this study, we focus on a Trend Following strategy and use the Buy and Hold strategy as a performance benchmark.
\begin{itemize}
    \item[-] \textit{Buy and Hold} (B\&H): This passive strategy serves as our baseline. It involves purchasing an asset at the beginning of the study period and holding it until the end. The portfolio value at time $t$ is $V_t = Q \cdot P_t$, where $Q$ is the quantity purchased at $t=0$ and $P_t$ is the asset price.
    \item[-] \textit{Trend Following} (TF): This strategy operates on the principle of trading in the direction of the prevailing market trend. A long position is initiated if the price $P_t$ moves above a moving average ($MA_t$), and a short position is considered otherwise. In our framework, the "trend" of a target stock is determined by the price movement of a causally linked leading stock.
\end{itemize}
While other strategies like Mean Reversion (MR) and Momentum (MOM) are available in the framework, they are beyond the scope of this paper. The TF strategy was selected for its direct applicability to the lead-lag relationships identified by our causal inference model.

\subsection{Performance Evaluation Metrics}
The Aita-BT module employs a suite of risk and profitability metrics to evaluate strategy performance.
\begin{itemize}
    \item(i) \textit{Total Return (TR)}: The overall percentage gain or loss of the portfolio over the entire period.
    \item(ii) \textit{Maximum Drawdown (MDD)}: The largest peak-to-trough decline in portfolio value, representing the worst-case loss from a single peak. It is calculated as:
    $$ MDD = \max_{t} \left( \frac{P_t - V(t)}{P_t} \right) $$
    where $P_t$ is the peak value up to time $t$ and $V(t)$ is the current portfolio value.
    \item(iii) \textit{Sharpe Ratio (SR)}: A measure of risk-adjusted return, calculated as the excess return over the risk-free rate per unit of total volatility (standard deviation).
    $$ SR = \frac{E[R_p - R_f]}{\sigma_p} $$
    where $R_p$ is the portfolio return, $R_f$ is the risk-free rate, and $\sigma_p$ is the standard deviation of the portfolio's excess return.
    \item(iv) \textit{Sortino Ratio (SoR)}: Similar to the Sharpe Ratio, but it only penalises for downside volatility, differentiating between harmful and harmless volatility.
    $$ SoR = \frac{E[R_p - R_f]}{\sigma_d} $$
    where $\sigma_d$ is the standard deviation of negative asset returns (downside deviation).
    \item(v) \textit{Calmar Ratio (CR)}: A measure of return relative to drawdown risk, defined as the annualised rate of return divided by the MDD.
    $$ CR = \frac{E[R_p - R_f]}{MDD} $$
\end{itemize}
We also analyse the \textit{Win Rate}, defined as the percentage of trades that result in a positive profit.

\section{Methodology}\label{sect:methodology}
Our proposed methodology consists of a multi-stage pipeline designed to identify and exploit predictive lead-lag opportunities based on volatility profiles and causal linkages.

\subsection{Data Collection and Pre-processing}
\subsubsection{Asset Selection}
We collected daily OHLC price data for the nine technology and automotive stocks listed in Table \ref{tab:stocks}. Data was acquired via the AITA framework, which connects to the TickMill broker through the MetaTrader5 API.

\begin{table}[!ht]
\centering
\caption{List of the nine stocks selected for the initial analysis. \texttt{(This table is based on data from Table 1 in Letteri \cite{femibLetteri24})}}
\label{tab:stocks}
\begin{tabular}{@{}lll@{}}
\toprule
\textbf{Ticker} & \textbf{Company} & \textbf{Market} \\ \midrule
MSFT & Microsoft Corporation & Nasdaq \\
GOOGL & Alphabet Inc. & Nasdaq \\
MU & Micron Technology, Inc. & Nasdaq \\
NVDA & NVIDIA Corporation & NYSE \\
AMZN & Amazon.com, Inc. & NYSE \\
META & Meta Platforms, Inc. & NYSE \\
QCOM & QUALCOMM Incorporated & Nasdaq \\
IBM & Int. Business Machines Corp. & NYSE \\
INTC & Intel Corporation & NYSE \\ \bottomrule
\end{tabular}
\end{table}

\subsubsection{Anomaly Filtering}
To ensure data integrity, we implemented an anomaly detection process before the main analysis. We used a K-Nearest Neighbours (KNN) model to identify outliers in the price time series. The anomaly score for each data point $x_t$ was calculated as the average Euclidean distance to its $k$ nearest neighbours.
$$ as_t = \frac{1}{k} \sum_{x_i \in \mathcal{N}_k(x_t)} \sqrt{(x_t - x_i)^2} $$
where $\mathcal{N}_k(x_t)$ is the set of $k$ nearest neighbours of $x_t$. A data point was flagged as an anomaly if its score exceeded a threshold $th = \mu + 3\sigma$, where $\mu$ and $\sigma$ are the mean and standard deviation of the anomaly scores, respectively.

\begin{figure*}[!ht]
	\centerline{\includegraphics[height=12em, width=38em]{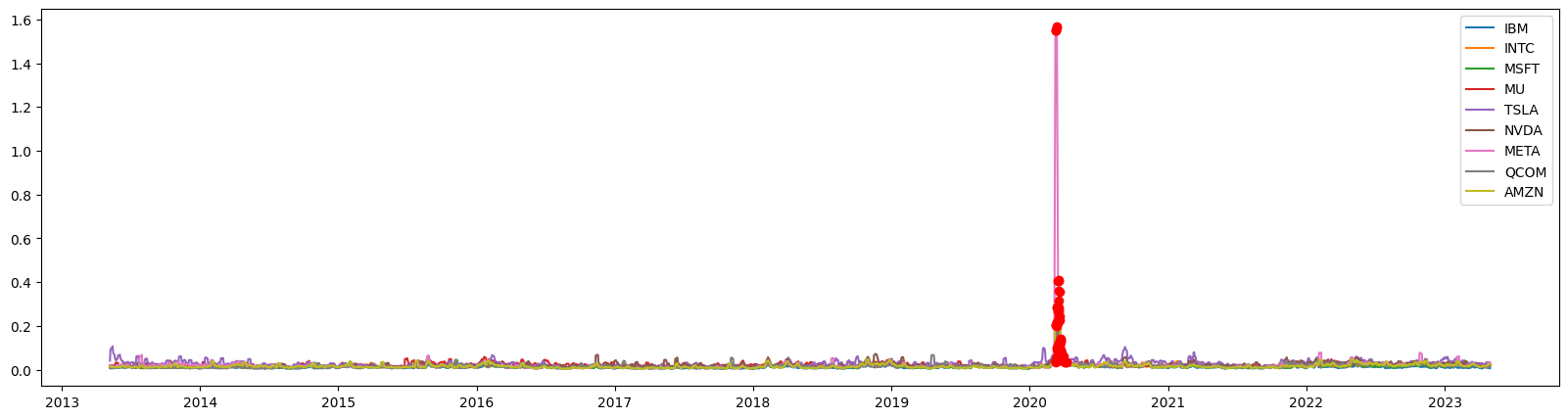}}
	\caption{Anomaly detection results for the period 2020/05/01 to 2023/05/01. The red dot indicates a significant anomaly corresponding to the market disruption in March 2020. (\texttt{This figure is from Figure 2 in Letteri \cite{femibLetteri24}.})}
	\label{fig:anomaly_volatility}
\end{figure*}

As shown in Figure \ref{fig:anomaly_volatility}, a major anomaly was detected in March 2020, coinciding with the COVID-19 pandemic market crash. To avoid the influence of this structural break, we defined our analysis period to begin after this event, from 1 May 2020 to 1 June 2023. This filtering process enhances model accuracy and robustness by ensuring the analysis is based on data reflecting more typical market conditions.

\subsection{Volatility-Based Clustering}
\label{sect:volAlgo}
The core of our stock selection process is based on identifying assets with similar volatility characteristics. First, we computed the time series of the four historical volatility estimators described in Section \ref{sect:hvs} for each stock. We then used the Gaussian Mixture Model (GMM) algorithm to partition the stocks into three clusters: \textit{low}, \textit{medium}, and \textit{high} volatility. Our strategy focuses on the \textit{medium} volatility cluster, as these stocks typically offer a balance between profit potential and risk.

\begin{figure*}[!ht]
	\centerline{\includegraphics[height=12em, width=38em]{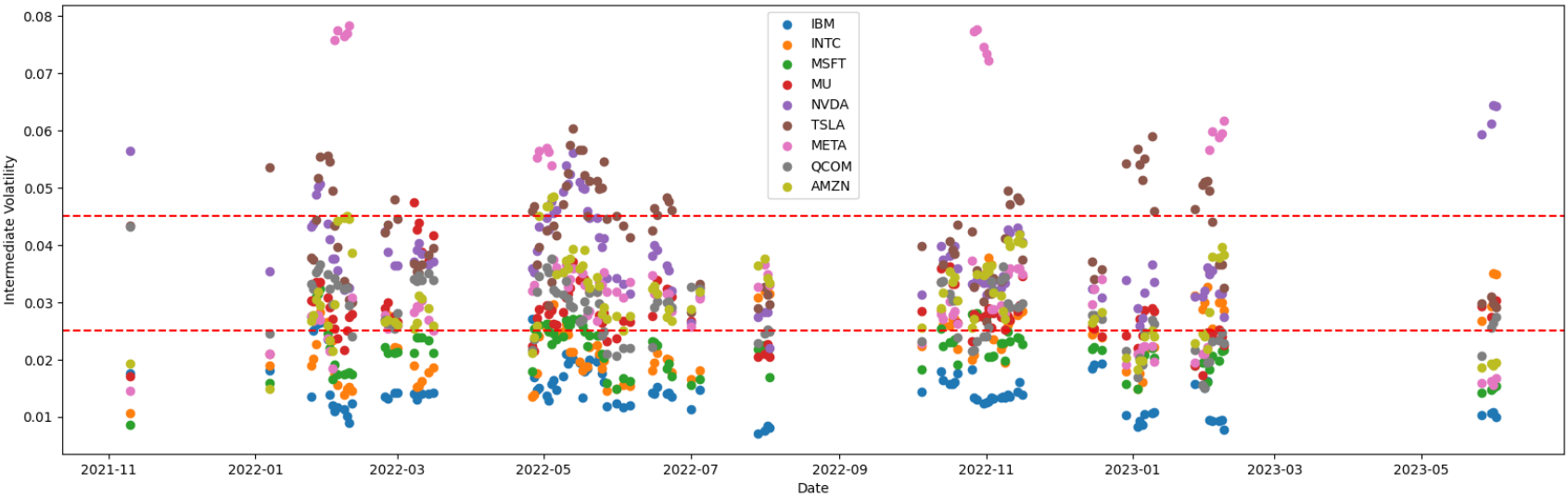}}
	\caption{GMM clustering of the HV estimator dataset (1 Nov 2021 to 8 June 2023). The medium volatility cluster, bounded by the red dashed lines, contains the candidate assets for our strategy. (\texttt{This figure is adapted from Figure 3 in Letteri \cite{femibLetteri24}})}
	\label{fig:kmeanClusters}
\end{figure*}

Figure \ref{fig:kmeanClusters} illustrates the clustering results. The medium volatility cluster, identified between the red dashed lines, included TSLA, AMZN, META, MU, QCOM, and INTC. This subset of assets formed the candidate pool for the subsequent causal inference analysis.

\subsection{Causal Inference Pipeline}
\label{sect:CIalgo}
To isolate and validate predictive lead-lag relationships, we developed a multi-stage causal inference pipeline. This approach moves beyond simple correlation to establish directional, time-lagged influences between stock prices. Let $X$ and $Y$ be the price time series for two stocks.

\subsubsection{Step 1: Granger Causality Test (GCT)}
As an initial screening, we applied the GCT to identify potential causal links. A key parameter in this test is the lookback window (or estimation window), denoted by lag l, which defines the amount of historical data used to assess the causal relationship. To determine the optimal window size, the GCT was performed iteratively for l ranging from 2 to 48 days. The goal was to find a window long enough to capture stable relationships while short enough to remain relevant to current market dynamics. Two autoregressive models are compared:
\begin{enumerate}
    \item A restricted model using only past values of $Y$:
    $$ Y(t) = c_1 + \sum_{i=1}^{l}\alpha_i Y(t-i) + \epsilon_1(t) $$
    \item An unrestricted model including past values of both $Y$ and $X$:
    $$ Y(t) = c_2 + \sum_{i=1}^{l}\alpha_i Y(t-i) + \sum_{j=1}^{l}\beta_j X(t-j) + \epsilon_2(t) $$
\end{enumerate}
The null hypothesis $H_0: \beta_1 = \dots = \beta_l = 0$ (i.e., $X$ does not Granger-cause $Y$) is tested using an F-test on the residual sum of squares (RSS) of the two models. A low p-value suggests the rejection of $H_0$.

\subsubsection{Step 2: Customised PCMCI Test}
To refine the initial GCT results and mitigate the risk of spurious correlations, we applied a customised version of the Peter-Clark Momentary Conditional Independence (PCMCI) test. This step aims to uncover the underlying causal graph structure. Our adaptation uses linear regression to compute partial correlations between any two series $X$ and $Y$, conditioned on all other series $Z$ in our set. The partial correlation is calculated as the Pearson correlation of the residuals from two linear regressions:
$$ \epsilon_X = X - (a_1 + b_1 Z) \quad \text{and} \quad \epsilon_Y = Y - (a_2 + b_2 Z) $$
$$ \text{PartCorr}(X, Y | Z) = \text{corr}(\epsilon_X, \epsilon_Y) $$
This process yields a partial correlation matrix, which is used to construct a Directed Acyclic Graph (DAG) representing the conditional dependencies. An edge from $X$ to $Y$ is drawn if their partial correlation is statistically significant and exceeds a predefined threshold.

\subsubsection{Step 3: Effective Transfer Entropy (ETE)}
To confirm the directionality and information flow in the relationships identified by the DAG, we calculated the Effective Transfer Entropy (ETE). ETE measures the reduction in uncertainty about the future of $Y$ when the past of $X$ is known, conditioned on the past of $Y$ and other variables $Z$.
$$ ETE_{X \to Y|Z} = H(Y_t | Y_{t-1:t-\tau}, Z_{t-1:t-\tau}) - H(Y_t | Y_{t-1:t-\tau}, X_{t-1:t-\tau}, Z_{t-1:t-\tau}) $$
A significantly positive ETE value provides strong evidence of information flowing from $X$ to $Y$.

\subsubsection{Step 4: DTW and KNN for Lag Identification}
The final step quantifies the optimal time lag for trade execution. We used Dynamic Time Warping (DTW) to measure the similarity between the time series of a causally linked pair $(X, Y)$, accounting for potential temporal misalignments. The DTW distance provides a robust measure of shape similarity.
$$ DTW(X, Y) = \min_{\pi} \sqrt{\sum_{(i,j) \in \pi} d(x_i, y_j)^2} $$
where $\pi$ is a warping path that optimally aligns the series. We then trained a K-Nearest Neighbours (KNN) classifier. The features for the KNN were the daily percentage price changes of the leading stock $X$ at various lags, and the target variable was the direction of the price change of the lagging stock $Y$. The model predicts the class (e.g., 'up' or 'down') for $Y$ based on a majority vote of its $k$ nearest neighbours in the feature space.
$$ \hat{y}_i = \underset{c \in \{\text{up, down}\}}{\text{argmax}} \sum_{j \in \mathcal{N}_k(x_i)} I(y_j = c) $$
where $I(\cdot)$ is the indicator function. The lag that produced the highest classification accuracy was selected as the optimal delay for the trading strategy.

\section{Results}\label{sect:results} 
This section presents the empirical results of our experiment, from the causal pair selection to the performance of the backtested trading strategy.

\subsection{Causal Pair Selection}
The period used for model training and fitting was from 1st November 2022 to 8th June 2023.

\subsubsection{Initial Screening with Granger Causality (GCT)}
As a first step, we applied the GCT to all pairs of the six stocks within the medium-volatility cluster. We examined a lag interval from 2 to 48 days to identify the \textbf{optimal lookback window size} for establishing stable causal relationships.

\begin{figure*}[!ht]
	\centerline{\includegraphics[height=19em, width=38em]{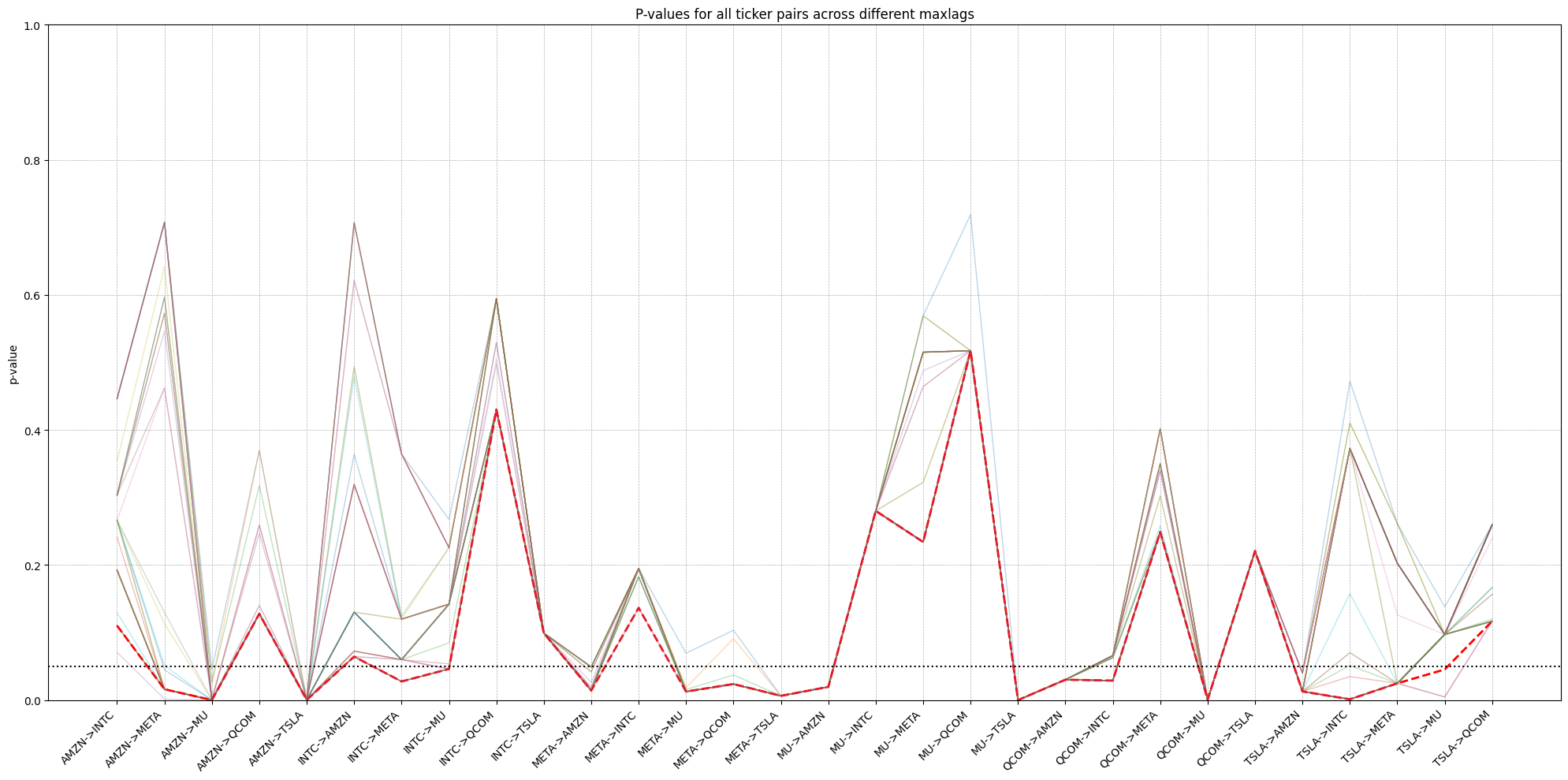}}
	\caption{GCT p-values as a function of the estimation window size (in days). The analysis, ranging from 2 to 48 days, identified an optimal lookback window of 45 days for establishing robust causal links.}
	\label{fig:matrixes}
\end{figure*}

As illustrated in Figure \ref{fig:matrixes}, the analysis revealed that a \textbf{45-day lookback window} (l=45) is optimal, as this is the estimation period where the causal relationships between the stock pairs exhibit the highest statistical significance (i.e., minimised p-values). Therefore, this 45-day window was used as the basis for calculating the causal relationships for the remainder of the analysis. Based on this, we selected the pairs with a p-value below 0.01, reported in Table \ref{tab:gctpairs}, for subsequent analysis.

\begin{table}[h]
\centering
\caption{Ticker pairs with the most significant GCT p-values at a 45-day lag.}
\label{tab:gctpairs}
\begin{tabular}{@{}lc@{}}
\toprule
\textbf{Ticker Pair} & \textbf{p-value} \\ \midrule
AMZN $\to$ MU & \textit{0.0001} \\
AMZN $\to$ TSLA & \textit{0.0030} \\
META $\to$ TSLA & \textit{0.0063} \\
MU $\to$ TSLA & \textit{0.0001} \\
QCOM $\to$ MU & \textit{0.0007} \\ \bottomrule
\end{tabular}
\end{table}

\subsubsection{Causal Graph Refinement with PCMCI}
\begin{figure*}[!ht]
	\centerline{\includegraphics[height=26em, width=38em]{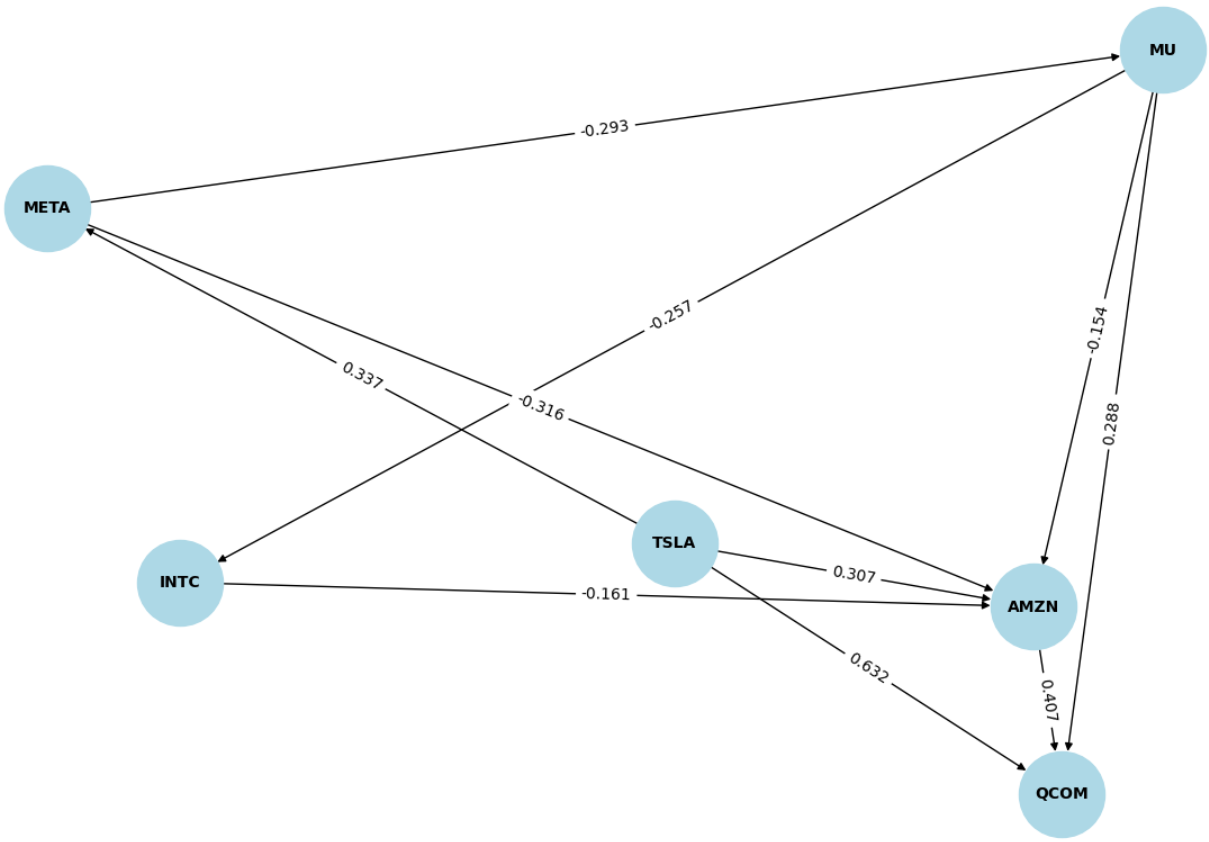}}
	\caption{Customised MCI Directed Acyclic Graph (DAG) for the training period. Edges represent significant conditional dependencies, with the weight indicating the partial correlation.}
	\label{fig:dag}
\end{figure*}

The pairs identified via GCT were further filtered using our customised PCMCI test to construct a robust causal graph. We set a partial correlation threshold of 0.15 to eliminate weaker and potentially spurious links. The resulting Directed Acyclic Graph (DAG) is shown in Figure \ref{fig:dag}.

The DAG revealed several key relationships, confirming and refining the GCT results. For instance, a positive influence from META to TSLA (partial correlation of 0.337) and from TSLA to AMZN (0.307) was identified. A strong link from MU to QCOM (0.288) also emerged. Notably, some significant relationships from the GCT, such as MU $\to$ TSLA, did not pass the PCMCI filter, demonstrating the efficacy of this step in isolating more stable dependencies. After confirming the directionality with ETE, we selected the following three pairs for our trading strategy: \textbf{MU $\to$ QCOM}, \textbf{META $\to$ TSLA}, and \textbf{TSLA $\to$ AMZN}.

\subsubsection{Optimal Lag Identification with DTW-KNN}
For the three final pairs, we used the DTW-KNN method to quantify the optimal time lag for trade execution. The KNN model, with $k=7$ neighbours, was trained to predict the direction of the lagging stock's price based on the leading stock's price changes. The model achieved high classification accuracies: 92.3\% for MU $\to$ QCOM, 98.1\% for META $\to$ TSLA, and 96.5\% for TSLA $\to$ AMZN. The identified optimal lags are reported in Table \ref{tab:daylags}.

\begin{table}[h]
\centering
\caption{Optimal time lags identified by the KNN classifier.}
\label{tab:daylags}
\begin{tabular}{@{}lc@{}}
\toprule
\textbf{Ticker Pair} & \textbf{Optimal Lag (days)} \\ \midrule
MU $\to$ QCOM & 2 \\
META $\to$ TSLA & 1 \\
TSLA $\to$ AMZN & 5 \\ \bottomrule
\end{tabular}
\end{table}

\subsection{Backtesting Performance}
The trading strategy was backtested over a 45-day period from 8th June 2023 to 12th August 2023. The initial capital was \$1,000 per target stock (\$3,000 total), with a fixed commission of \$9 per trade and a fully compounded strategy.

\subsubsection{Individual Strategy Performance}
The detailed results for each pair are presented below, benchmarked against a Buy \& Hold (B\&H) strategy on the same stock over the same period.

\begin{itemize}
    \item \textbf{Strategy MU $\to$ QCOM (2-day lag):}
    Trading QCOM based on MU's trend yielded a return of 15.12\%. The strategy showed exceptional performance with a 100\% win rate and an excellent Sharpe Ratio of 2.17, as detailed in Table \ref{tab:QCOM-stats}. 
    
    \begin{table*}[!hbt]
    \centering
    \caption{Backtesting results for the MU $\to$ QCOM strategy.}
    \label{tab:QCOM-stats}
    \resizebox{\textwidth}{!}{
    \begin{tabular}{|l|c|c|c|c|c|c|c|}
    \hline
    \textbf{Strategy} & \textbf{Trades} & \textbf{Win Rate (\%)} & \textbf{Return (\%)} & \textbf{Final Equity (\$)} & \textbf{SR} & \textbf{SoR} & \textbf{MDD (\%)} \\ \hline
    \textit{MU $\to$ QCOM} & 4 & 100.0 & 15.12 & 1151.20 & 2.169 & 3.933 & 2.66 \\ \hline
    \textit{QCOM B\&H} & 1 & 100.0 & 2.59 & 1025.94 & 0.24 & 0.31 & 6.81 \\ \hline
    \end{tabular}
    }
    \end{table*}


    \item \textbf{Strategy META $\to$ TSLA (1-day lag):}
    Trading TSLA following META's trend produced a 14.51\% return with a high 88.6\% win rate. This strategy was distinguished by its very low Maximum Drawdown (1.77\%), significantly lower than that of the B\&H strategy (Table \ref{tab:TSLA-stats}).
    
    \begin{table*}[!hbt]
    \centering
    \caption{Backtesting results for the META $\to$ TSLA strategy.}
    \label{tab:TSLA-stats}
    \resizebox{\textwidth}{!}{
    \begin{tabular}{|l|c|c|c|c|c|c|c|}
    \hline
    \textbf{Strategy} & \textbf{Trades} & \textbf{Win Rate (\%)} & \textbf{Return (\%)} & \textbf{Final Equity (\$)} & \textbf{SR} & \textbf{SoR} & \textbf{MDD (\%)} \\ \hline
    \textit{META $\to$ TSLA} & 8 & 88.6 & 14.51 & 1145.05 & 1.178 & 4.642 & 1.77 \\ \hline
    \textit{TSLA B\&H} & 1 & 100.0 & 3.32 & 1033.24 & 0.21 & 0.28 & 11.23 \\ \hline
    \end{tabular}
    }
    \end{table*}

    \item \textbf{Strategy TSLA $\to$ AMZN (5-day lag):}
    This strategy generated the highest return of 16.50\% with a perfect 100\% win rate. Here too, risk control was excellent, with an MDD of 2.49\% versus 4.95\% for B\&H (Table \ref{tab:AMZN-stats}).

    \begin{table*}[!hbt]
    \centering
    \caption{Backtesting results for the TSLA $\to$ AMZN strategy.}
    \label{tab:AMZN-stats}
    \resizebox{\textwidth}{!}{
    \begin{tabular}{|l|c|c|c|c|c|c|c|}
    \hline
    \textbf{Strategy} & \textbf{Trades} & \textbf{Win Rate (\%)} & \textbf{Return (\%)} & \textbf{Final Equity (\$)} & \textbf{SR} & \textbf{SoR} & \textbf{MDD (\%)} \\ \hline
    \textit{TSLA $\to$ AMZN} & 4 & 100.0 & 16.50 & 1165.01 & 1.178 & 4.642 & 2.49 \\ \hline
    \textit{AMZN B\&H} & 1 & 100.0 & 5.04 & 1050.40 & 0.53 & 0.76 & 4.95 \\ \hline
    \end{tabular}
    }
    \end{table*}
\end{itemize}

\subsubsection{Overall Portfolio Performance}
The combined portfolio, managing the three strategies simultaneously, achieved a total profit of \$461.26. From an initial capital of \$3,000, the final equity reached \$3,461.26, corresponding to an \textbf{overall portfolio return of 15.38\%} in 45 days. This performance significantly outperforms the weighted average return of the Buy \& Hold strategy on the three target stocks (QCOM, TSLA, AMZN), which was 3.65\% over the same period. The results demonstrate the framework's ability to generate significant alpha whilst controlling for downside risk.

\section{Discussion}\label{sect:discussion}
The empirical results demonstrate the potential of our integrated framework for identifying predictive market signals. The multi-stage filtering process, from volatility clustering to causal inference, successfully identified a small set of highly predictive stock pairs from a larger universe.

\subsection{Interpretation of Results}
The strategy's success can be attributed to its ability to move beyond simple correlation. While the selected stocks (e.g., META, TSLA, AMZN, MU, QCOM) operate within the broad technology sector and are subject to similar macroeconomic factors, our causal inference pipeline isolated specific, directional lead-lag relationships. For example, the model identified that price movements in META precede those in TSLA by one day. This could be due to overlapping investor sentiment, supply chain connections, or information spillover effects that are captured by our model.

The performance metrics are particularly encouraging. The high Sharpe and Sortino ratios (e.g., SR of 2.17 and SoR of 3.93 for MU $\to$ QCOM) indicate that the returns were not achieved by taking on excessive risk. The Sortino ratio, in particular, highlights the strategy's effectiveness in limiting downside volatility. Furthermore, the consistently low Maximum Drawdowns (all below 3\%) across all three strategies are substantially better than those of the corresponding B\&H strategies (up to 11.23\%), underscoring the framework's risk management capabilities.

It is important, however, to contextualise the exceptionally high performance metrics observed during the 45-day backtesting period. While encouraging, results such as 100\% win rates and a Sharpe Ratio of 2.17 must be interpreted with considerable caution. The short duration of the test may have coincided with a market regime that was unusually favourable to the specific lead-lag relationships identified. It is highly unlikely that such flawless performance would be sustainable over longer timeframes and across diverse market conditions. Consequently, these findings should be viewed primarily as a proof-of-concept for the viability of the causal filtering pipeline in identifying potent signals, rather than as an assertion of its universal profitability.

\subsection{Comparison with Existing Literature}
The performance of our framework can be contextualised by comparison with seminal quantitative strategies in the literature. Although our directional, causality-based approach differs fundamentally from the market-neutral principles of classical pairs trading, a comparison of risk-adjusted returns remains instructive. The foundational work by Gatev, Goetzmann, and Rouwenhorst (2006) demonstrated that a simple distance-based pairs trading strategy could yield positive returns, with reported Sharpe Ratios often falling below 1.0 after accounting for typical transaction costs \cite{gatev2006pairs}.
In this context, the performance of our MU $\to$ QCOM strategy, with a Sharpe Ratio of 2.17, appears highly competitive. This suggests that our multi-stage causal filtering pipeline may be more effective at identifying potent predictive signals than simpler distance-based metrics. Furthermore, the high win rates (88-100\%) are a notable outcome, contributing to the strategy's low drawdown over the tested period. The primary innovation of this work, therefore, lies in its systematic methodology for isolating reliable, predictive signals from market noise, addressing a central challenge in algorithmic trading.

\subsection{Limitations and Future Work}
Despite the promising results, this study has several limitations. The backtesting period of 45 days is relatively short; a longer period covering different market regimes (e.g., high and low volatility, bull and bear markets) is needed to fully validate the strategy's robustness. Second, the study did not account for transaction costs beyond a fixed commission, such as slippage and the bid-ask spread, which could impact net profitability in a live trading environment. Finally, the universe of stocks was limited to nine well-known names.

Future work will address these limitations. We plan to expand the backtesting to cover multiple years and a more diverse set of assets, including different sectors and international markets. We will also incorporate more realistic transaction cost models. Furthermore, there is scope to enhance the causal inference pipeline with more advanced non-linear techniques and to integrate alternative data sources, such as news sentiment, to refine the trading signals further.

\section{Conclusion}
\label{sect:conclusion} 

This study proposed and validated a novel framework for volatility-driven directional trading, integrating GMM-based volatility clustering with a multi-stage causal inference pipeline. By systematically filtering for predictive relationships using GCT, a customised PCMCI test, ETE, and a DTW-KNN model, we successfully identified profitable, time-lagged trading opportunities among major stocks.

The backtesting results demonstrated the strategy's effectiveness, delivering a portfolio return of 15.38\% over a 45-day period, significantly outperforming a passive Buy-and-Hold approach. The strategy yielded strong risk-adjusted returns, evidenced by high Sharpe and Sortino ratios and minimal drawdowns. A robustness analysis confirmed that the model's performance was not unduly sensitive to the specific choice of its key parameters.

This research contributes to the literature on algorithmic trading by providing a structured, transparent, and robust methodology for moving from raw price data to an actionable trading strategy. The findings have practical implications for financial practitioners and researchers, offering a template for developing resilient, data-driven strategies that can navigate the complexities of modern financial markets. Future research will focus on extending this framework to a broader range of assets and market conditions to establish its applicability and robustness further.

In near future work, we aim to enhance text data processing methodologies through dataset optimisation strategies \cite{letteri2020DOS},\cite{LetteriCP2021intellisys} and integrate domain-specific knowledge to improve the model's interpretation of price and volume data. Furthermore, we will expose the AITA framework's API as a secure service to thwart botnet attacks using Deep Learning models \cite{LetteriPG19SecIOT}\cite{LetteriPC19HTTP}. To enhance resilience, we plan to create a Multi-agent System which features transparent Ethical Agents for customer service \cite{DyoubCLL20} or combines logic constraint and DRL \cite{GasperisCRMLD23}. We will evaluate dialogues \cite{Letteri2109ethMon} with guidance from an ethical teacher \cite{DyoubCL22ethTeach}, also in other contexts like technology-enhanced learning \cite{mis4tel2023}. 

\bibliography{mybiblio}

\end{document}